\documentclass[aps,preprint,floatfix,superscriptaddress,amsmath,showpacs,amssymb]{revtex4}

\usepackage{amssymb}
\usepackage{epsfig}
\usepackage{dcolumn}
\usepackage{bm}

\begin{document}

\title{Ca impurity in small mixed $^4$He-$^3$He clusters}
\author{R. Guardiola}
\affiliation{IFIC (CSIC-Universidad de Valencia), Apartado Postal 22085, E-46.071-Valencia, Spain}
\author{J. Navarro}
\affiliation{IFIC (CSIC-Universidad de Valencia), Apartado Postal 22085, E-46.071-Valencia, Spain}
\author{D. Mateo}
\affiliation{Departament E.C.M., Facultat de F\'{\i}sica, and IN$^2$UB, 
Universitat de Barcelona. Diagonal 647, E-08028 Barcelona, Spain}
\author{M. Barranco}
\affiliation{Departament E.C.M., Facultat de F\'{\i}sica, and IN$^2$UB, 
Universitat de Barcelona. Diagonal 647, E-08028 Barcelona, Spain}

\date{\today}

\begin{abstract}
The structure of small mixed helium clusters doped with one calcium atom has been 
determined within the diffusion Monte Carlo framework. 
The results show that the calcium atom sits at the $^4$He-$^3$He interface.
This is in agreement with previous studies, both experimental and theoretical, 
performed for large clusters. A comparison between the results obtained for the 
largest cluster we have considered
for each isotope shows a clear tendency of the Ca atom to reside in a deep dimple 
at the surface of the cluster for $^4$He clusters, and to become fully solvated for $^3$He clusters.
We have calculated the absorption spectrum of Ca around the $4s4p \leftarrow 4s^2$ 
transition and have found that it is blue-shifted from that of the free-atom transition 
by an amount that depends on the size and composition of the cluster.
\end{abstract}


\maketitle

\section{Introduction}
Helium clusters are weakly bound quantum systems as a consequence of the small atomic mass 
and the weak van der Waals interaction between helium atoms. There are several recent 
papers~\cite{toen04,barr06,stie06,cho06,tigg07,sza08}
reviewing the most relevant aspects of the physics and chemistry of these remarkable systems, 
to which we refer the interested reader.

It was found about twenty years ago that helium clusters could pick up closed shell molecules 
singly and that these molecules were located in the interior of the clusters~\cite{sche90,sche93}. 
It is now well established that helium clusters may virtually capture any kind of atoms and molecules,
most of which are located in their bulk. Alkali atoms are the sole known exception, as they remain 
on the surface of the cluster upon their capture.
Electronic spectroscopy allows to obtain information on the cluster structure, and in particular on 
the position of embedded impurities in the cluster, via the shift and width of the electronic
transitions~\cite{stie01,stie06}. 

Heavy alkaline-earth atoms but magnesium exhibit a different behavior for each helium isotope, in 
contrast with most impurities, which solvate similarly in both. 
They are solvated in $^3$He but not in $^4$He clusters, in which case they reside in dimples on 
the cluster surface. 
Although still debated, the magnesium atom is most likely in a very delocalized state within the 
bulk of the $^4$He cluster. 

Clusters formed by a mixture of both helium isotopes are particularly interesting, as they are 
made of bosons and fermions with different mass interacting through the same potential. It is 
known that in the bulk, the excess in kinetic energy pushes the $^3$He atoms to the liquid 
surface~\cite{andr66}, an effect that also shows up in mixed droplets, see {\em e.g.} Ref.~\onlinecite{barr97}.
It has been experimentally found that, depending on the size and composition of the drop, 
Ca atoms reside in bubbles at or near the $^4$He-$^3$He interface~\cite{bune09}. This scenario has been 
confirmed by density functional (DF) calculations~\cite{bune09,mate09}.
Heavy alkaline-earth atoms thus may offer a unique opportunity to study the interface of 
isotopically mixed helium clusters~\cite{hern07}.

In this paper we address the study of small mixed $^4$He-$^3$He clusters doped with one Ca atom.  
Our calculations are based on diffusion Monte Carlo (DMC) methods using current realistic 
interatomic interactions. In particular, we have calculated density distribution functions and have estimated the atomic 
shifts and widths for Ca attached to a cluster, for several combinations of the number of isotopes in it.
Results for some $^4$He and $^3$He clusters doped with Ca are also presented to complete the discussion. 
It is worth stressing that, while $^4$He clusters doped with different impurities have been 
thoroughly studied, theoretical works on doped $^3$He clusters are
scarce~\cite{barl03,lopez04,delara06,delara09a,delara09b}
apart from those carried out within the DF approach.
The paper is organized as follows. In Sec.~\ref{method} we give some details about the DMC 
calculations. In Sec.~\ref{results} we present our results, and in Sec.~\ref{conclusion}
we present a brief summary.

\section{Method}
\label{method}
The DMC description is based on a variational or importance sampling wave function.
We have used a rather simple form, which contains the basic required properties. It is a 
generalization of the trial function adopted in our previous studies on pristine mixed helium 
clusters~\cite{guar00a,guar02}. It is written as a product of seven terms
\begin{equation}
\Psi({\bf{\cal R}}) = \Psi_{44} \, \Psi_{33} \Psi_{34} \Psi_{4{\rm Ca}} \Psi_{3{\rm Ca}}
D_{\uparrow} D_{\downarrow} 
\label{function}
\end{equation}
consisting of a Jastrow factor for each pair of different atoms, plus the spin-up and -down
Slater determinants required to satisfy the Pauli exclusion principle for $^3$He fermions.
Here $\{ {\bf{\cal R}} \}$ represents the set of $3(N_4+N_3+1)$ coordinates of the atoms forming the cluster. 
Any of the Jastrow terms has the generic form
\begin{equation}
\Psi_{MN} = \prod_{i  \neq j} \exp\left( -\frac{1}{2} \left[\frac{b_{MN}}{r_{ij}}\right]^{\nu_{MN}} -
\alpha_{MN} r_{ij} \right)  \; ,
\label{bosons}
\end{equation}
where indices $i,j$ run over the corresponding type of atom, and includes
a short-range repulsion term with parameters $b_{MN}$ and $\nu_{MN}$ associated to it,
and a long-range confining term with corresponding parameter $\alpha _{MN}$.
The so defined $\Psi_{MN}$ function is explicitly symmetric under the exchange of particles.

The antisymmetry required for $^3$He fermions is incorporated in the Slater determinants
$D_{\uparrow}$ and $D_{\downarrow}$, related to the spin-up and -down fermions.
These Slater determinants are of primary relevance because they define the set of spin-up 
and -down nodal surfaces which strongly constrain the DMC algorithm.
As in our previous works~\cite{guar00a,guar02} we have assumed a shell-model like structure, 
taking the single-particle orbitals as harmonic polynomials of the fermionic cartesian coordinates, 
thus guaranteeing that the resulting Slater determinants are translationally invariant. 
Moreover, we have always assumed a filling scheme in which the total spin is minimum, either 0 or 1/2, 
respectively, for $N_3$ even or odd. Also, the so called Feynman-Cohen back-flow~\cite{FC} has been 
incorporated into the scheme by substituting 
\begin{equation}
{\bf r}_i \rightarrow {\tilde{\bf r}}_i =  {\bf r}_i + \sum_{i \ne j} \eta(r_{ij})
({\bf r}_i - {\bf r}_j) 
\label{backflow}
\end{equation}
in the Slater determinants~\cite{LK}. For the backflow function $\eta(r)$ we choose the medium-range 
form used in Ref.~\onlinecite{pand86}, namely
$\eta(r) = \lambda/r^3$, with the same value of $\lambda= 5$~\AA$^3$.

The form of the short-range repulsion term was introduced long ago by 
McMillan~\cite{mcmi65}  for describing the homogeneous liquid $^4$He using
a 12-6 Lennard-Jones interaction. The values of the two parameters $\nu$ and $b$ are fixed by the short-range behavior of a pair of atoms. 
In our calculations we have employed the He-He Aziz HFD-B(HE)~\cite{aziz87} 
potential, and that of Hinde~\cite{hind03} for the $X ^1\Sigma$ Ca-He interaction, as  parametrized 
in Ref.~\onlinecite{mate09}. 
For all cluster sizes we have used the following values:
 $\nu_{MN}=5.2$~\AA, $b_{44}=2.95$~\AA,  $b_{33}=2.85$~\AA, $b_{34}= 2.90$~\AA, 
as in previous works on pristine and isotopically mixed helium clusters, and 
$b_{4{\rm Ca}}=3.25$~\AA, and $b_{3{\rm Ca}}=3.00$~\AA. 
Thus, the trial or importance sampling wave function contains only five free parameters, namely 
$\alpha_{44}$, $\alpha_{33}$, $\alpha_{34}$, $\alpha_{4{\rm Ca}}$, and $\alpha_{3{\rm Ca}}$, which 
have been determined by minimizing the expectation value of the Hamiltonian.

For the DMC algorithm we have used the  short-time Green function approximation
~\cite{ande80,reyn82} with an $O(\tau^3)$ form~\cite{vrbi86}.
It is worth mentioning that this approximate Green function satisfies the microreversibility condition 
(detailed balance condition). Moreover, the random process was constrained so as not to traverse the 
nodal surfaces, using the so called fixed node approximation.

\section{Results}
\label{results}
\subsection{Structure and energetics of Ca@$^4$He$_{N_4}$+$^3$He$_{N_3}$ clusters}

\begin{figure}[h!]
\epsfig{file=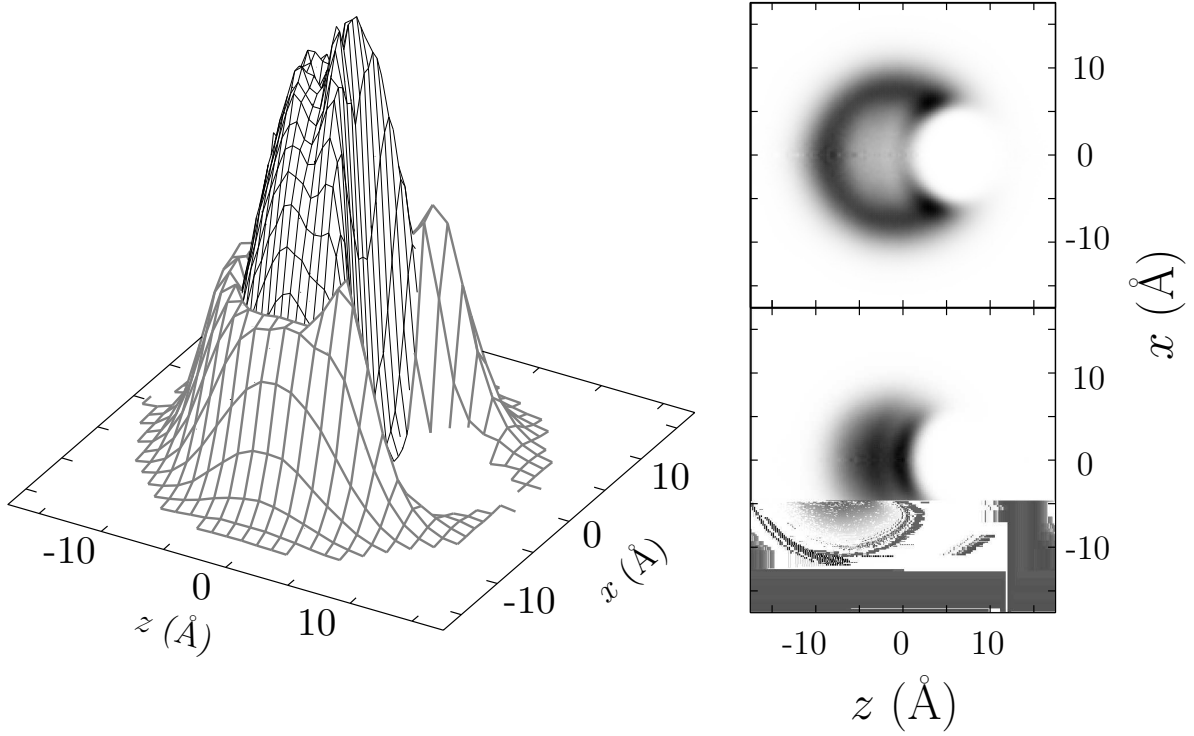,scale=0.9}
\caption{Density distributions in the $y=0$ plane for the Ca@$^4$He$_{20}$+$^3$He$_{20}$ 
cluster. Left: 3D representation. Right: grey scale plot of $^3$He (top) and $^4$He (bottom) 
densities. The darker the region, the higher the helium density.}
\label{den2020}
\end{figure}

\begin{figure}[h!]
\epsfig{file=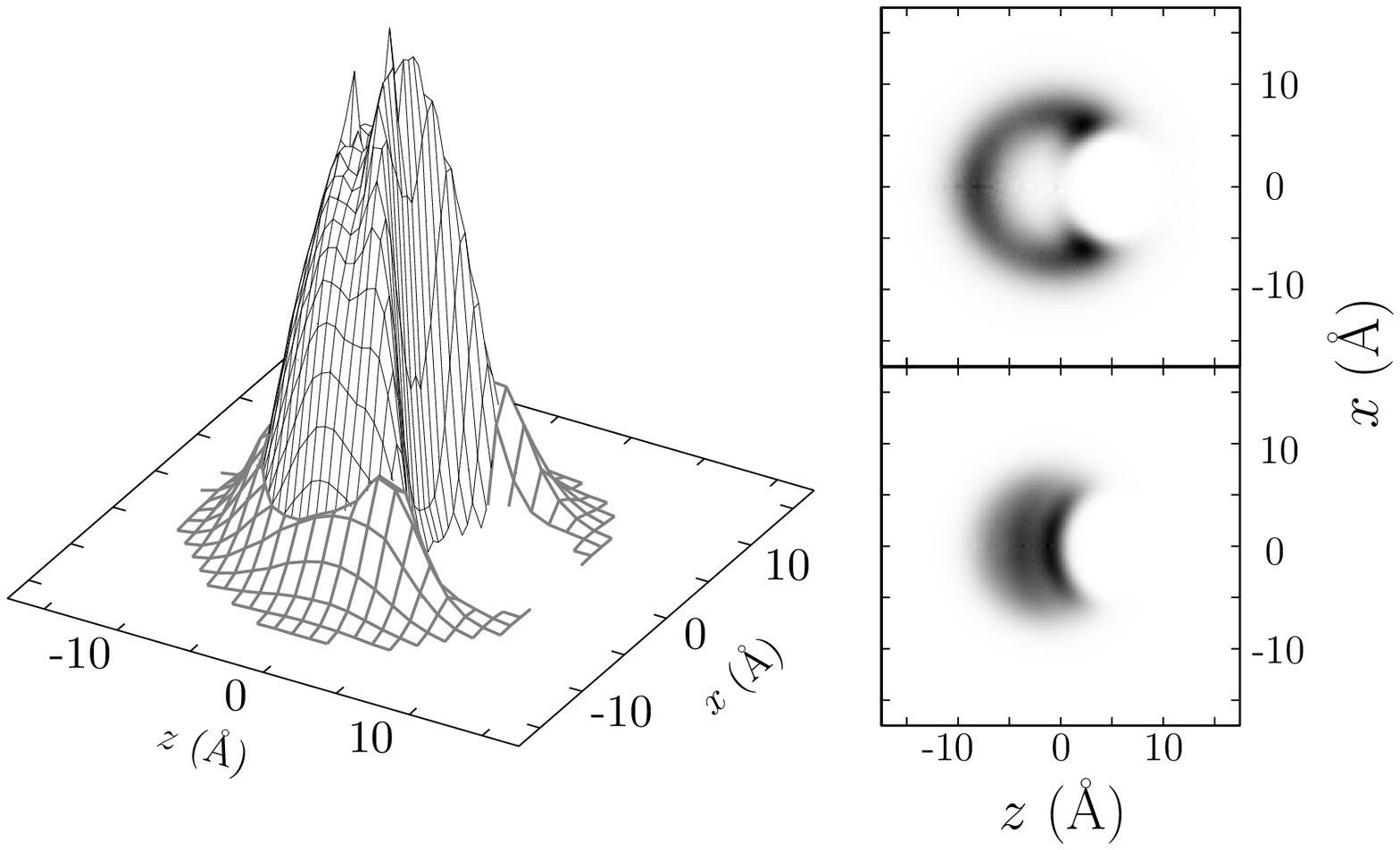,scale=0.9}
\caption{Same as Fig.~\ref{den2020} for $N_4=20$, $N_3=8$.}
\label{den2008}
\end{figure}

\begin{figure}[h!]
\epsfig{file=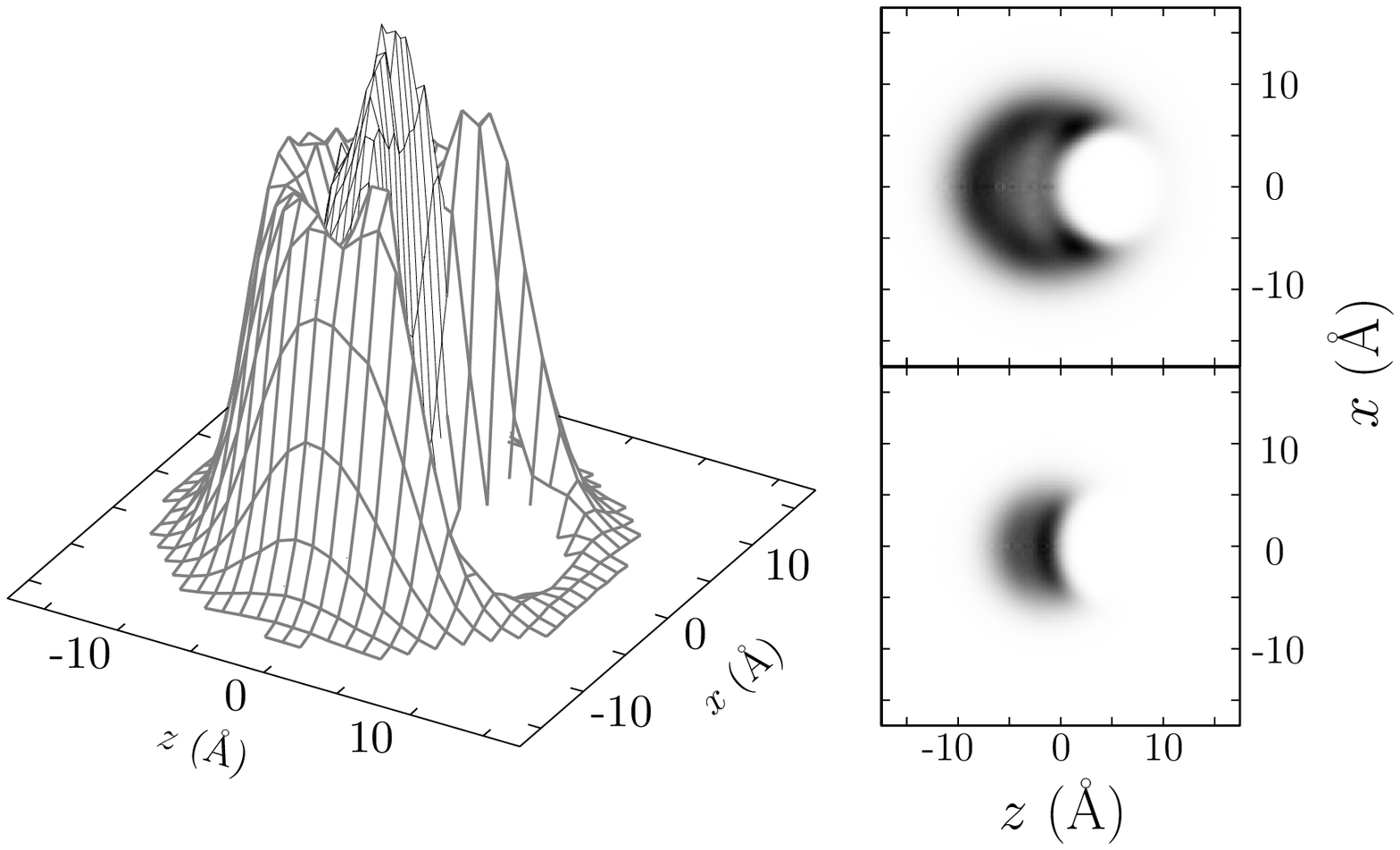,scale=0.9}
\caption{Same as Fig.~\ref{den2020} for $N_4=8$, $N_3=20$.}
\label{den0820}
\end{figure}

\begin{figure}[h!]
\epsfig{file=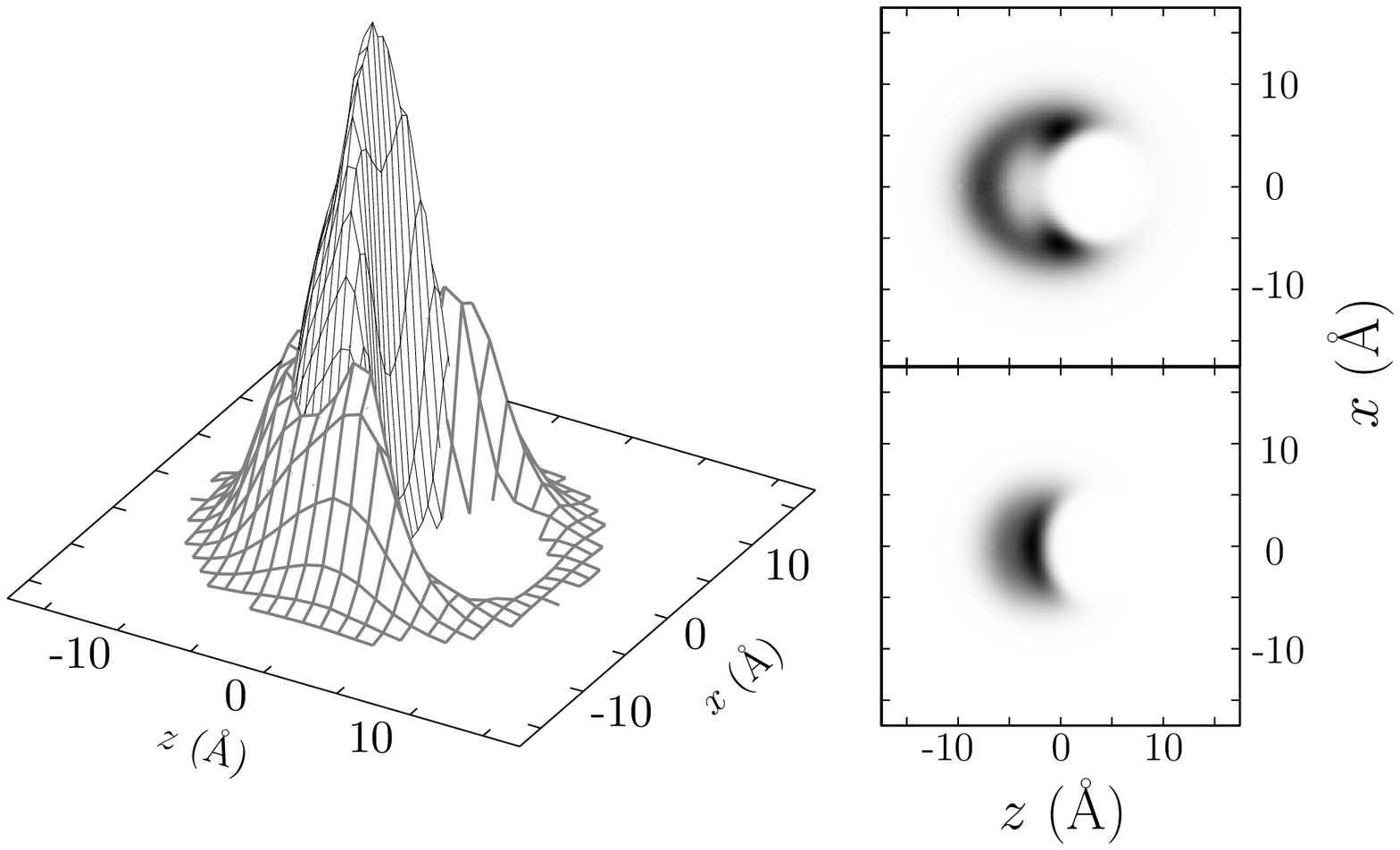,scale=0.9}
\caption{Same as Fig.~\ref{den2020} for $N_4=8$, $N_3=8$.}
\label{den0808}
\end{figure}

In Figures \ref{den2020}-\ref{den0808} are plotted the densities in the $y=0$ plane of the 
$(N_4,N_3) =$ (20,20), (20,8), (8,20), and (8,8) clusters.
After a simulation running for a long thermalization time we have stored a large number of walkers 
(typically 10$^6$) to get a mixed estimator of the probability densities. A series of angular 
rotations have been performed on each walker to place the Ca atom on the $z$-axis,
defined as the line joining the impurity and the cluster center-of-mass. 
Afterwards, a projection on the $y=0$ plane has been performed. 
The resulting plots thus exhibit axial symmetry. 

These clusters display an already well known structure, with a core of $^4$He atoms surrounded by a 
shell of $^3$He atoms. Interestingly, in spite of the small number of helium atoms, 
the tendency of Ca to reside at the $^4$He-$^3$He interface is clearly visible.
Note also the peak in the $^4$He densities near to Ca, and those appearing in the $^3$He shell: the 
one near the impurity is reminiscent of the solvation shell that would fully develop in a larger cluster 
made of either kind of atoms. We attribute the $^3$He peak distant from the impurity to the tendency 
of the $^3$He atoms to reside, whenever possible, far from the impurity~\cite{mate09b} because the 
Ca-He interaction is weaker than the He-He one. Within DF theory, a qualitatively similar structure 
has been found for clusters with $N_4=50$ and $N_3=$ 18, 32, 50, and 68~\cite{mate09}.

The calculated ground state energies of the clusters, as well as the solvation energy of the dopant, 
defined as
\begin{equation}
S({\rm Ca}) = E({\rm Ca}@^4{\rm He}_{N_4}+^3{\rm He}_{N_3}) - E(^4{\rm He}_{N_4}+^3{\rm He}_{N_3})
\;\; ,
\label{solvation}
\end{equation}
are given in Table~\ref{ener34}.  We mention that Elhiyani and Lewerenz~\cite{elhi09} have carried 
out calculations for Ca@$^4$He$_{N_4}$ clusters using the same He-He interaction than ours,
and a Ca-He pair potential they have obtained at the CCSD(T) level of accuracy, 
as Hinde's one~\cite{hind03}. Their Ca solvation energies for clusters with $N_4=$ 20 and 40 
are -12.29~(1) and -16.55~(6)~K,  respectively, in good agreement with ours as displayed in Table~\ref{ener34}.

\begin{figure}[h!]
\epsfig{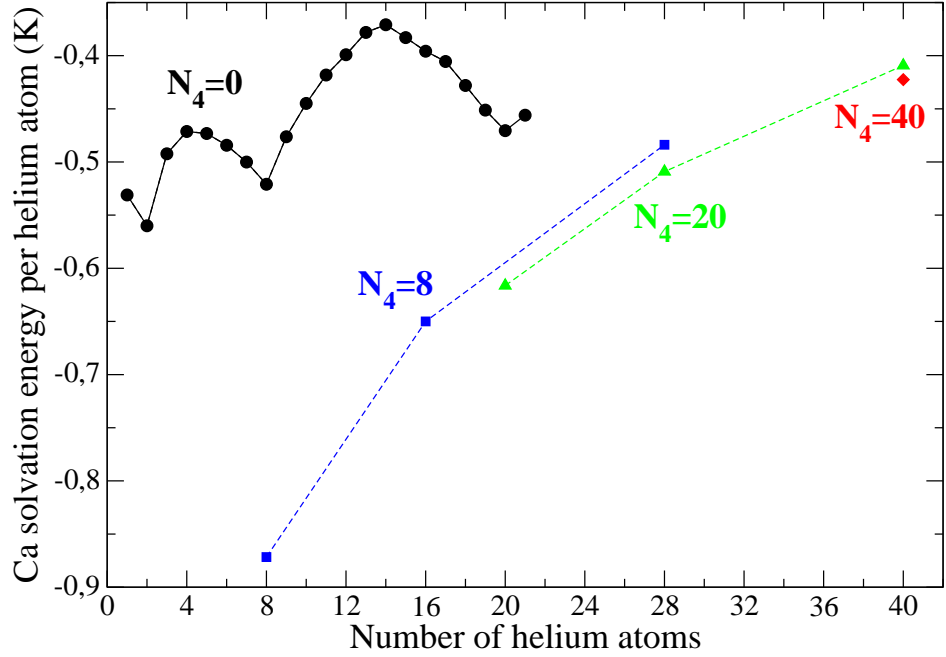}
\caption{(color online) Solvation energy of Ca per helium atom for
several Ca@$^4$He$_{N_4}$+$^3$He$_{N_3}$ clusters, and total energy per
helium atom of Ca@$^3$He$_{N_3}$ as a function of the total number of
helium atoms in the cluster.}
\label{solvat34}
\end{figure}

In Fig.~\ref{solvat34} are plotted the Ca solvation energies per helium atom as a function of the total 
number of helium atoms in the cluster. Note the conspicuous shell oscillations for $N_4=0$. We discuss them below.

\subsection{The case of Ca@$^3$He$_{N_3}$ clusters}
 
The study of the lightest $^3$He clusters has a great interest, as they are a challenge for theoretical
methods due to their very small binding energy. In particular, pure $^3$He clusters are bound only for 
sizes greater than $N_3 \sim 30$ atoms~\cite{barr97b,guar00a,sola06}, and small mixed $^4$He-$^3$He 
clusters present instability regions as a function of $N_4$ and 
$N_3$~\cite{guar02,bres03,kornilov}. 
It has been shown that the presence of one  $^4$He atom is enough to trigger the binding of twenty
$^3$He atoms~\cite{guar02}, showing the importance of the zero-point motion in the stability of small 
helium clusters.

\begin{figure}[h!]
\epsfig{file=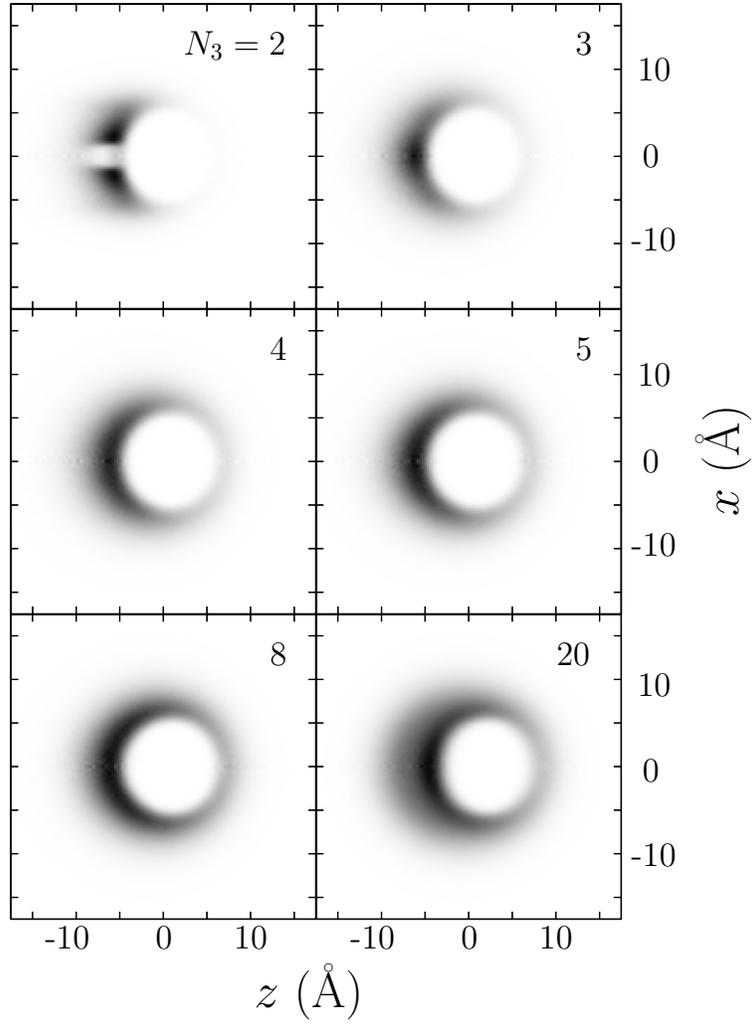,scale=0.9}
\caption
{Grey scale plot of the
density distributions in the $y=0$ plane for
Ca@$^3$He$_{N_3}$ clusters with $N_3$= 2, 3, 4, 5, 8, and 20.}
\label{CONTcahe3}
\end{figure}

The Ca-He interaction potential is weaker than the He-He one, and its minimum lies at a larger distance. 
Nevertheless, due to the different zero-point energy of Ca and He atoms, we have found that the presence 
of one single Ca atom causes that clusters made of any number of atoms are bound,
irrespective of their bosonic or fermionic character. This is an obvious result for $^4$He clusters, but it is
not for $^3$He clusters. We give in Table~\ref{ener3} the total energy of Ca@$^3$He$_{N_3}$ clusters, and 
have already represented them, divided by $N_3$, in Fig.~\ref{solvat34} ($N_4=0$ results) because 
as the pristine $^3$He clusters are not bound, it makes sense to plot these energies together with 
the solvation energy of Ca discussed before. It is interesting to see the sharp minima at 
$N_3 = 2$, 8, and 20. These numbers correspond to the first shell closures of the three-dimensional 
spherical harmonic oscillator, in spite that the helium density distributions are far from being 
spherical, especially for small
$N_3$ values (see Fig.~\ref{CONTcahe3}). We have no plausible explanation for this finding. However, 
it is worth recalling that in cylindrical symmetry \cite{note}, the first two minima also arise from the
$1 \sigma$ and $1 \sigma, 2 \sigma, 1 \pi$ shell closures corresponding to this geometry, 
see {\em e.g.}, Ref.~\onlinecite{mate09}.

\begin{figure}[ht]
\epsfig{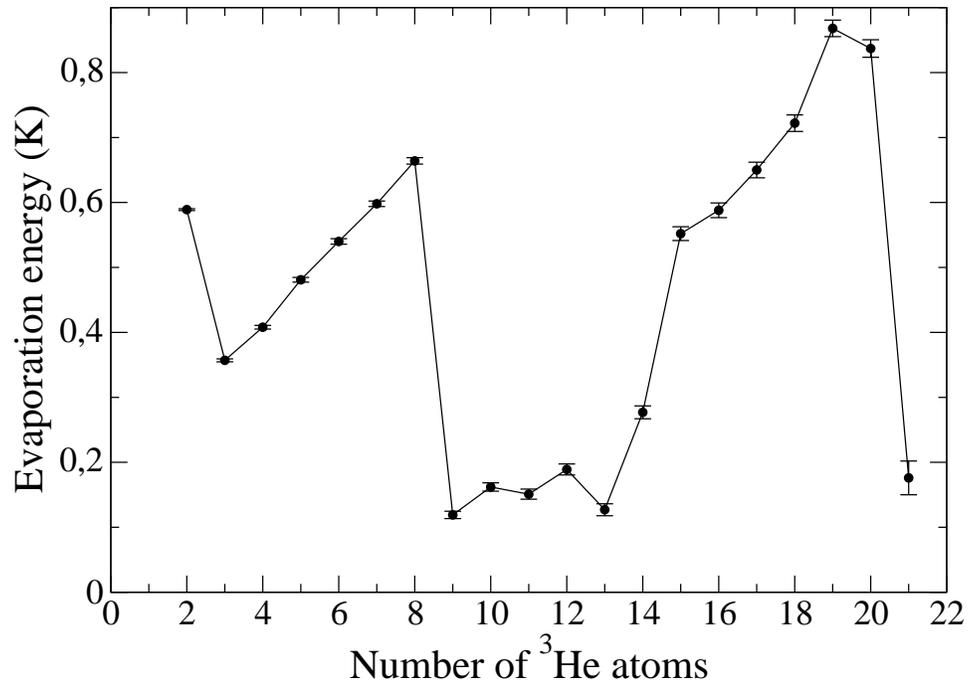} 
\caption{
Evaporation energies of $^3$He atoms in Ca@$^3$He$_{N_3}$ clusters
for the indicated N$_3$ values. }
\label{vapor}
\end{figure}

For Ca@$^3$He$_{N_3}$ clusters we may also calculate the differences
\begin{equation}
\Delta(N_3)=E({\rm Ca}@^3{\rm He}_{N_3})-E({\rm Ca}@^3{\rm He}_{N_3-1})
\end{equation}
than can be interpreted as the negative of the He evaporation 
energies~\cite{mell05}. We have plotted  $-\Delta(N_3)$ in
Fig. \ref{vapor}. The large drops in the evaporation energy
after $N_3 = 2, 8$, and 20 correlate well with the ``magicity'' 
of these values, i.e., with shell closures.
In the $N_3 = 9-20$ range, the evaporation energy
displays some structure, with a weak local maximum at $N_3=12$.

\begin{figure}[h!]
\epsfig{file=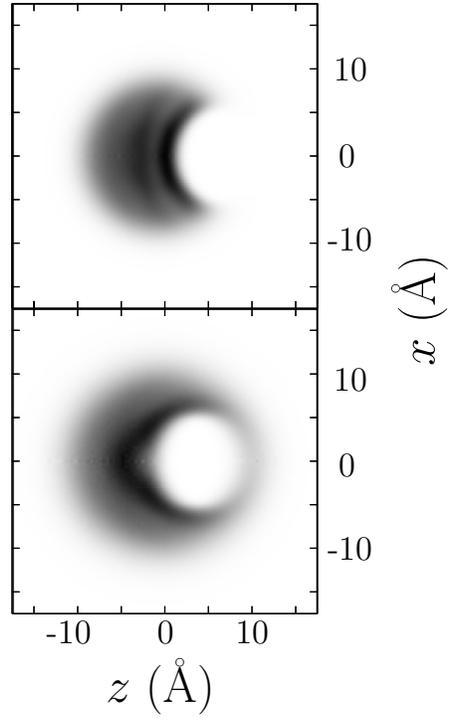,scale=0.9}
\caption{Grey scale plot of the density distributions in the
$y=0$ plane for
Ca@$^4$He$_{40}$ and Ca@$^3$He$_{40}$.}
\label{CONTcahe4cahe3}
\end{figure}

As representative examples, the density distributions in the $y=0$ plane for the
Ca@$^3$He$_{N_3}$ clusters with $N_3$= 2, 3, 4, 5, 8, and 20 are plotted in 
Fig.~\ref{CONTcahe3}. It can be clearly seen the tendency of the Ca atom to 
become solvated in $^3$He clusters.
It is also interesting to compare the morphology of $^4$He and $^3$He clusters doped with Ca.
To this end, we have plotted in Fig.~\ref{CONTcahe4cahe3} the density distributions 
corresponding to helium clusters made of 40 atoms doped with it. It can be seen that 
Ca is in a dimple state on the surface of the $^4$He cluster, whereas it moves towards 
the bulk of the cluster in the case of
$^3$He. In spite of this bulk location, the Ca@$^3$He$_{40}$ cluster is not spherically symmetric. 
This also happens for clusters doped with other weakly attractive impurities, such as Mg in small $^4$He
clusters~\cite{mell05}. The reason is the tendency of the system to have a helium density 
as close as possible to its liquid value in order to maximize its binding. This is easier 
to archieve in an
off-center location for small clusters, even for $N_3=2$. In this case the two $^3$He atoms 
are slightly apart as a consequence of their large zero-point motion, not fully compensated 
by the attraction of the Ca impurity. This is the reason of the very low He density
in a cylindrical region surrounding the $z$-axis, see the corresponding panel in
Fig. \ref{CONTcahe3}, which is absent in the other clusters.
Eventually, when the number of $^3$He atoms increases, 
the center-of-masses of the helium moiety and the impurity nearly coincide~\cite{hern07}.

\subsection{Absorption spectrum of Ca in$^4$He$_{N_4}$+$^3$He$_{N_3}$ clusters}

The DMC calculation provides us with a set $\{ {\bf {\cal R}} \}$ of walkers indicating the 
instantaneous position of each atom in the cluster. From these walkers we have calculated 
the density distribution
functions plotted in the previous subsections. Here we use them to obtain the absorption spectrum 
of the $4s4p \, ^1P_1 \leftarrow 4s^2 \, ^1S_0$ transition in Ca.

Electronic spectroscopy is a powerful tool to disclose the structure of impurities in helium 
clusters, since the shift and width of the electronic transitions is very sensitive to the dopant 
environment~\cite{stie06,tigg07}. Lax method~\cite{lax52} has been since long time ago the standard 
way to determine the absorption spectrum of a dopant atom in helium clusters. It makes use of 
the Franck-Condon principle within a semiclassical approach. It has been adapted and used by several 
authors to analyze the absorption spectrum of {\em e.g.}, lithium in solid H$_2$~\cite{chen96}, 
and of different atoms in helium clusters~\cite{naka01,mell02,hern08,hern09}. The particular case of Ca
atoms we are interested in has been addressed within DF in pure and mixed helium clusters~\cite{hern08,bune09}.
Lax method is usually applied in conjunction with the diatomics-in-molecules approach~\cite{ell63}, 
in which the atom-cluster complex is treated as a diatomic molecule, the helium moiety playing the 
role of the other atom.

We have calculate the line shape of the electronic absorptium transition in Ca as
\begin{equation}
I(\omega)  \propto \int {\rm d}{\bf {\cal R}} |\Psi_{gs}({\bf {\cal R}})|^2 
\delta( \omega+V_{gs}({\bf {\cal R}})-V_{ex}({\bf {\cal R}}) ) \;\; ,
\end{equation}
where $\{ \bf{\cal R} \}$ refers to the positions of the atoms, and $V_{gs}$ and $V_{ex}$ 
are, respectively, the ground and excited states potential energy surfaces.
We have used the Ca-He $X^1\Sigma$ interaction of Ref.~\onlinecite{hind03} for the ground state, and 
the $^1\Pi$ and $^1\Sigma$ potentials of Ref.~\onlinecite{hern08} for the excited states. In short, 
for a given value of $\omega$, a $3\times 3$ matrix has to be diagonalized to determine the three 
components of the absorption line, each one arising from a different potential energy surface, 
i.e., eigenvalue of the excited energy matrix, see Refs.~\onlinecite{hern08,hern09b} for the details.

\begin{figure}[h!]
\epsfig{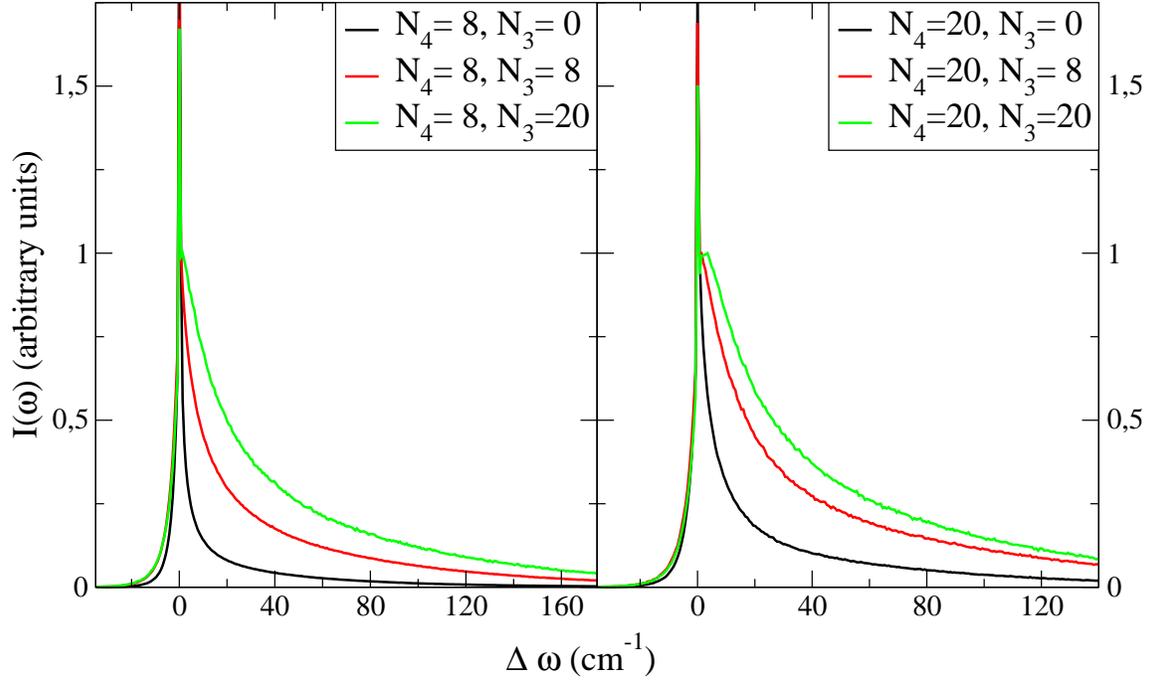} 
\caption{(color online) 
Absorption spectrum of Ca in mixed clusters with
$N_4=8$ (left panel) and $N_4=20$
(right panel), and three $N_3$ values.}
\label{abs1}
\end{figure}

In Fig.~\ref{abs1} is plotted the absorption spectrum of Ca for several combinations of $N_4$ and $N_3$ 
values. The energies are referred to the free Ca atom value (23650 cm$^{-1}$). 
For so few helium atoms, there is no appreciable shift, but an appreciable width increasing as $N_3$ 
does at fixed $N_4$.
A large enough number of walkers bear the information corresponding to a quasi-free Ca atom within the 
simulation volume. This is the reason of the narrow peak at $\omega=0$. A similar peak appears in the 
experiments, and it is due to the excitation of gas-phase Ca atoms present in the doping chamber
\cite{sti97}.

\begin{figure}[h!]
\epsfig{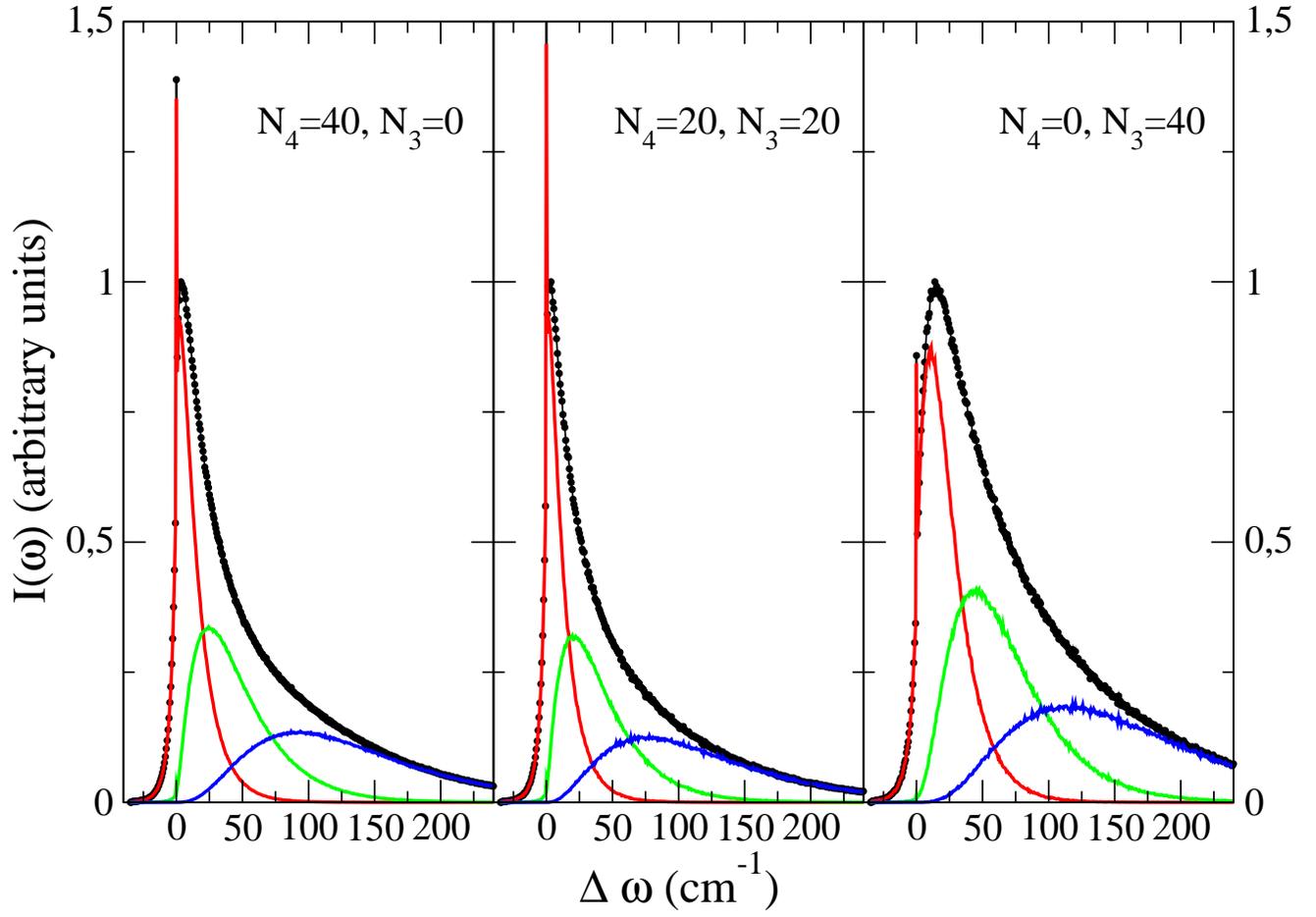}
\caption{(color online)
Absorption spectrum of Ca for three selected combinations
of helium atoms in clusters with $N_4+N_3=40$.}
\label{abs2}
\end{figure}

\begin{figure}[ht]
\epsfig{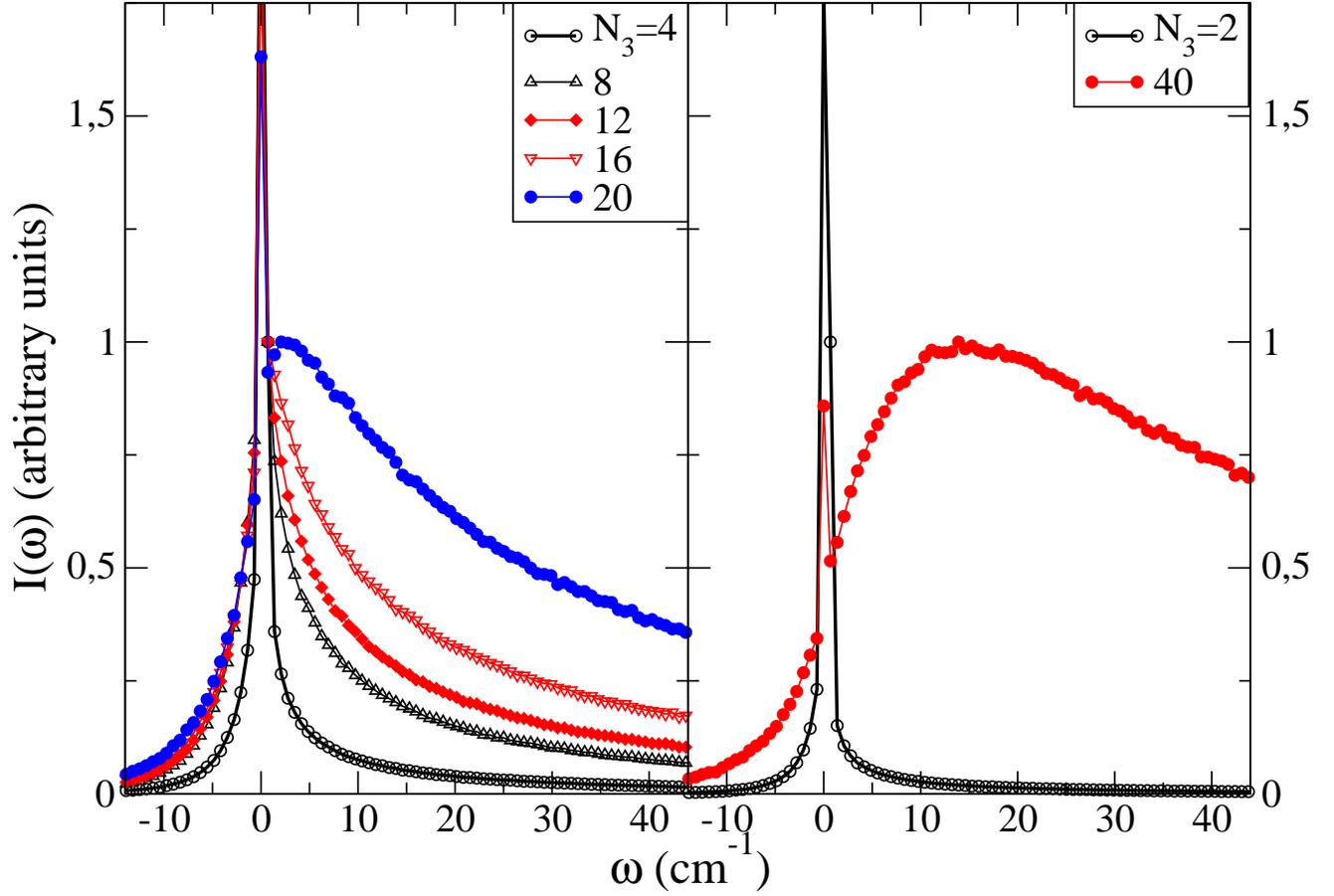} 
\caption{(color online)
Absorption spectrum of Ca in $^3$He$_{N_3}$ clusters
for the indicated number of atoms. For clarity, the
$N_3=2$ and 40 results have been drawn in a separate panel.}
\label{abs3}
\end{figure}

In Fig.~\ref{abs2} is plotted the absorption spectrum of Ca for three selected $(N_4, N_3)$ combinations 
with $N_4+N_3=40$. The contributions from the three potential energy surfaces are shown, 
and the long tail at high frequencies is identified as likely arising from the very repulsive contribution 
of the $^1\Sigma$ pair potential. The
shift in the (0, 40) case is clearly seen. Finally, in Fig.~\ref{abs3} we display the absorption spectrum
of Ca@$^3$He$_{N_3}$ for increasing values of $N_3$. Not surprisingly, the $N_3=2$ case
shows no appreciable shift and a small width. As $N_3$ increases, both the shift and width progressively do. 

\section{Summary}
\label{conclusion}

Within the diffusion Monte Carlo framework, we have studied the structure and energetics of small mixed
$^4$He-$^3$He clusters doped with Ca. We have found that a single Ca atom is able to produce bound 
clusters made of any number of helium atoms, irrespective of their fermionic or bosonic character. In 
the case of $^3$He clusters, this is a somewhat unexpected result, as the Ca-He interaction is weaker 
than the He-He one. We attribute it to a zero-point motion effect.

We have found that Ca resides in a deep dimple at the surface of $^4$He clusters, and in the bulk of 
$^3$He clusters, in agreement with experiments carried out for large helium 
clusters~\cite{bune09,sti97}.

In the case of mixed clusters, our calculations show the tendency of the impurity to sit at the 
$^4$He-$^3$He interface, although the small number of fermions that we can microscopically handle does not
allow us to make a clear distinction between the interface and the rest of the mixed cluster.
Morphologically, we have found that doped mixed clusters obtained within DF theory~\cite{mate09} 
are qualitatively similar to those obtained within the microscopic DMC approach.

Finally, we have used the DMC walkers to semiclassically obtain the absorption spectrum of Ca 
around the $4s4p \leftarrow 4s^2$ transition, and have found that it is slightly blue-shifted 
from that of free-atom transition, with a clear dependence on the size and composition of the 
mixed cluster in spite of the small number of helium atoms in the studied systems.
We want to mention that the excitation and emission spectra of Ca atoms implanted in
liquid $^4$He and $^3$He have been measured \cite{mori05}, and that their determination
in the case of liquid mixtures could also be carried out as a function of pressure and
composition. We are at present extending the method implemented in Ref. \cite{bune09} to
address this issue.

\acknowledgments

We thank Alberto Hernando and Mart\'{\i} Pi for useful discussions and comments,
and Mohamed Elhiyani and Marius Lewerenz for providing
us with their results prior to publication.
This work has been supported by Grants FIS2007-60133 and FIS2008-00421
from DGI, Spain (FEDER), and  2009SGR1289 from Generalitat de
Catalunya.

\newpage

\begin{table}[htb]
\caption{Ground state energies (in units of K) of mixed helium clusters,
pristine and Ca doped, and solvation energy of Ca for several combinations
of $N_4$ and $N_3$.}
\begin{center}
\begin{tabular}{rr|c|c|c}
\hline
$N_4 $ &$N_3$ & $E(^4$He$_{N_4}$+$^3$He$_{N_3})$ &
$E($Ca@$^4$He$_{N_4}$+$^3$He$_{N_3})$ & $S$(Ca)  \\
\hline
 8   &    0 &   -5.086 (3) & -12.059 (4) & -6.973 (5) \\
  8   &   8 & -11.884 (14) & -22.286 (9) & -10.402 (17) \\
  8   & 20 & -20.514 (16) & -34.061 (20) & -13.547 (22) \\
 20  &  0  & -33.427 (10) & -45.752 (16) & -12.325 (19) \\
 20  &  8  & -46.705 (20) & -60.959 (25) & -14.254 (32) \\
 20  & 20 & -62.265 (41) & -78.634 (50) & -16.369 (65) \\
 40  &  0  & -101.935 (25) & -118.836 (14) & -16.901 (29)  \\
\hline
\end{tabular}
\end{center}
\label{ener34}
\end{table}

\newpage 

\begin{table}[htb]
\caption{Ground state energies 
of Ca@$^3$He$_ {N_3}$ clusters}
\begin{center}
\begin{tabular}{rc|rc|rc}
\hline
$N_3$ & Energies (K) &  $N_3$ & Energies (K)  &  $N_3$ & Energies (K)  \\
\hline
 1   &  -0.531 (1) &  8   & -4.168 (4)  &  15   & -5.745 (8) \\   
 2   & -1.120 (1)  &  9   & -4.287 (4)  &   16   & -6.333 (8)   \\  
 3   & -1.477 (2)  & 10  &- 4.449 (5)  &   17   & -6.983 (9)   \\   
 4   & -1.885 (2)  &  11  & -4.600 (6)  & 18   & -7.705 (9)     \\ 
 5   & -2.366 (3)  &  12   & -4.789 (6) & 19   & -8.573 (9) \\   
 6   & -2.906 (3)  &  13   & -4.916 (7) & 20   & -9.410 (10)   \\   
 7   & -3.504 (3)  &    14   & -5.193 (7)   & 21 & -9.586 (24) \\
\hline
\end{tabular}
\end{center}
\label{ener3}
\end{table}

\newpage

\newpage

\newpage

\newpage

\newpage

\newpage

\newpage

\newpage

\newpage

\newpage

\newpage

\begin{thebibliography}{99}
\bibitem{note}
In axial geometry, the projection of the angular momentum on the
symmetry axis is the good quantum number, and single-particle
states with angular momentum $\pm l_z$ are degenerate. Thus,
taking into account the spin degeneracy, levels with $l_z=0$
are twofold degenerate, and those with $l_z \neq 0$ are fourfold
degenerate.
\end{thebibliography}
\end{document}